# Playing magic tricks to deep neural networks untangles human deception

Regina Zaghi-Lara[1], Miguel Ángel Gea[2], Jordi Camí[3], Luis M. Martínez[4], Alex Gomez-Marin[1,*]

[1] *Behavior of Organisms Lab, Instituto de Neurociencias CSIC-UMH, Alicante*
[2] *Teatro Encantado, Madrid*
[3] *Universitat Pompeu Fabra, Barcelona*
[4] *Visual Analogy Lab, Instituto de Neurociencias CSIC-UMH, Alicante*
[*] Correspondence*: agomezmarin@gmail.com*

**Abstract.** Magic is the art of producing in the spectator an illusion of impossibility. Although the scientific study of magic is in its infancy, the advent of recent tracking algorithms based on deep-learning allow now to quantify the skills of the magician in naturalistic conditions at unprecedented resolution and robustness. In this study, we deconstructed stage magic into purely motor maneuvers and trained an artificial neural network (DeepLabCut) to follow coins as a professional magician made them appear and disappear in a series of tricks. Rather than using AI as a mere tracking tool, we conceived it as an "artificial spectator". When the coins were not visible, the algorithm was trained to infer their location as a human spectator would (i.e. in the left fist). This created situations where the human was fooled while AI (as seen by a human) was not, and vice versa. Magic from the perspective of the machine reveals our own cognitive biases.

Magic is not the violation of the natural order of things, but the command of cognitive processes. This is why it is said that real magic is fake, while fake magic is real. After centuries of trial-and-error, magicians have empirically learned to hijack several mechanisms of attention, perception memory and instinctive decisions. Tricks take place less in the deck of cards than in the spectator's mind.

The insights that magic offers to psychology and cognitive science are not new [**1,2**], yet attempts for a science of magic are only recently flourishing [**3,4**]. For instance, the spectator's gaze can be measured with eye tracking technology to better understand the magician's misdirection [**5**] (**Figure 1A**). Yet, precise quantitative measures of the skills of the magician have been elusive.

DeepLabCut is a recently-developed open-source software based on supervised deep learning for video analysis by marker-less pose estimation with transfer learning [**6**]. It offers an unparalleled opportunity to measure human behavior in complex activities outside the lab, such as the magician's prestidigitation. Here, rather than using it as a mere tracking tool, we conceived it as an "artificial spectator": we can study how the magician's fingers move (**Figure 1B**), but also guess where the coins are as he makes them appear and disappear (**Figure 1C**). We propose a double inversion of perspective: from human to machine; from motor skills to perceptual inference.

We asked a professional magician to perform a series of sleight of hand coin tricks, which we specifically designed to be simple but effective enough to fool a human only via brief pure motor maneuvers (no narrative or verbal instructions, social cues, special effects or the use of gimmicks). We then conceived the magician's trick from the point of view of the machine (**Figure 1D**). We trained the algorithm to follow the position of the coins in each trick, were they visible or not. By labeling the coins in a few frames, a human trained the network to automatically label the rest of video, as explained in [**6**]. Although a sensible recommendation for tracking is not to label parts of interest that are occluded (to skip those frames by simply not labeling anywhere) here, instead, we trained the network also with images in which the coins were not visible: the human annotating frames clicked where she thought the coin was (i.e. in the left fist; under the right palm). The machine learned those human priors (**Figure 1E**).

The magic tricks performed, tracked and analyzed are shown in the **Supplementary Video**. We encourage the reader to watch them before reading further. They should all feel like magic to most naive spectators, but it is of course possible that some humans do not get fooled in some of them. This attests to the individuality and diversity of modes of perception that magicians know well and that they face every time they stand in front of an audience. Moreover, context is constitutive of the illusion of impossibility (it is not the same to see a brief magic clip on a tablet at home than a whole set in a theater with more people clapping and laughing). A magic trick should naturally happen once and at the peace imposed by the magician, rather than repeatedly, paused and slowed down by the spectator at will. Note that one can be "un-fooled" after being fooled —which permits greater insights into the phenomenon, as attempted here—, but not the opposite (one cannot make oneself a magic trick either). The "good" magic spectator is not the analytic observer who resists to be fooled, but the enchanted mind that voluntarily dwells into the experience of the impossible. As scientists, we can approach magic in both ways: experiencing first, to then deconstruct it.

The set consisted of five main brief tricks (plus two extra ones; see below), all related to cognitive illusions. Each was based on a different maneuvers that the magician performs on the coins but that the human fails to notice due to the naturality of the cover movement: throwing (trick #1), dropping (trick #2), dragging (trick #3), placing (trick #4), grasping (trick #5). In the **Supplementary Video**, right after the sequence of raw videos, we show them again with a red spot marking the machine's inference of coin location. Then, in **Figure 1F** we classify each trick depending on whether it would fool the human or the machine (the human-learned prior by the machine). During tricks #1 and #2 the coin is not visible during most of the sleight of hand for neither human or AI. Both infer it to be in the closed fist, where it is not. This shows that at least some cognitive tricks can in fact transfer from humans to machines. In trick #3 the coins are visible at times and not visible at other times, the effect being due to perceptual overload in the human spectator. In the "machine Umwelt", each hand is watched separately (actually, pixel by pixel) and so the trick vanishes. In trick #4 the coin is actually visible at the moment of the effect, but the human tends not to see it due to misdirection of attention. The machine has no problem detecting the fourth coin at the very moment it is visibly placed on the table. Speed (the machine analyzes frame after frame at "its own pace") and attention (the machine analyzes the whole image pixel by pixel, not choosing where to look at) are only an issue for humans. In trick #5 it should be visible which coins are grasped with each hand, but Gestalt symmetry principles confuse us making us assume each hand takes one. The machine has not learned such priors.

Our main goal was to provide a few cases corresponding to both left quadrants in **Figure 1F**, as just described. Yet, by asking the magician to perform magic differently from what he would, we can scarcely provide examples for the right quadrants. In trick #6 (corresponding to trick #1, done badly on purpose), one can actually see how the coin briefly flies from one hand to the other. In this case, the machine does not detect it, because it appears sideways, looking like a rod rather than a circle, which is what it has learned to see. And so it still believes the coin is in the left fist. Finally, in trick #7 (corresponding to trick #5, done slowly on purpose) one can more easily see how many coins each hand actually reaches. This quadrant corresponds to neither the machine or the human being fooled. Let us mention the interesting case (which we do not show here) in which the trick would be time-reversed. For the machine, detection is invariant if video frames are fed to it forward or backward in time. And so, if it were fooled forwards, it would also be fooled backwards. Since "no magic" is different from confusion or a mess, nothing of this sort would work for humans (except for potential "palindrome-like tricks").

Our work has several limitations. First, we did not train the network to label every frame as "fooled" versus "not fooled". The quadrants in **Figure 1F** (specially the orange one) reflect what the human from the perspective of the machine would claim with respect to each trick. Second, we trained with frames of the same video that was automatically labeled, rather than a separate one. Third, sometimes, if the coin is occluded and the hand posture not detected as the trained one (i.e. closed fist), coin inference can fail (i.e. the algorithm flagging the magician's eyes, or his jacket buttons). Fourth, we did not create any new deep learning algorithm, but used an existing one for a purpose it was not designed for. While originally used for digit tracking of mice during hand reaching (among other behaviors) [**6**], here we used DeepLabCut to study the perceptual effects of a sleight of hand. Fate, it seems, is not without a sense of irony.

In sum, merging magicians and machines can open unexplored ways to investigate human cognition, where the enchanted mind of the spectator, the analytic mind of the scientist and the artificial mind of a neural network meet. We have shown that it might be possible to target the development of adversarial cognitive tricks at humans by choosing AI models that match the human cognitive system as closely as possible. As the analogy between adversarial images [**7**] and adversarial tricks remains to be concretely drawn, our results are a prelude to further deployments of deep learning in the context of the demanding processes of a magic performance —a perspective to our knowledge not explored so far ("artificial illusionism"). AI still needs to be equipped with our familiar notions of causation [**8**]. And yet we can learn from what machines learn from us [**9**]. By trying to establish similarities between human and machine behavior [**10**], we will better appreciate our differences. The combination of the millenary art of magic with the latest advances in AI ("mAgIc") creates a new powerful mirror to look into ourselves.

*

**Methods.** Experimental procedures were approved by the "Oficina de Investigación Responsable" (OIR, UMH). We recorded the magician with GoPro cameras at his theater (Teatro Encantado). We used DeepLabCut 2.0 software for tracking, whose open source code can be found here: https://github.com/AlexEMG/DeepLabCut. For every tracked video, no refinement of labels was needed, but the original labeling (which varied across tricks; typically of the order of 5% of the total frame number) with a training of at least 200K iters per video.

**Supplementary information.** The data used in this study, compiled into raw and tracked magic tricks, can be found as a Supplementary Video here: https://youtu.be/KPizTPQz0tc

**Contributions**. Idea: AGM; conceptualization: LMM & AGM; experimental design: JC, LMM, & AGM; magic: MAG; labeling and analyses: RZL; figure, video and manuscript: AGM.

**Acknowledgements.** We thank Alexander Mathis & Mackenzie Mathis for creating and sharing DeepLabCut, and for useful comments. We acknowledge Streamline for figure icons.

**Funding.** This work was supported by the Spanish Ministry of Science (BFU-2015-74241-JIN and RYC-2017-23599 grants to AGM).

**Declaration**. The authors declare no competing financial interests.

# References

[**1**] Binet A (1894) La Psychologie de la Prestidigitation. Rev Deux Mondes. 125:903-22.
[**2**] Jastrow J (1896) Psychological notes upon sleight-of-hand experts. Science. 3:685-9.
[**3**] Macknik S, King M, Randi J, Robbins A, Teller, Thompson J, & Martinez-Conde S (2008) Attention and awareness in stage magic: turning tricks into research. Nature Reviews Neuroscience 9:871-9.
[**4**] Kuhn G (2019) Experiencing the Impossible. The Science of Magic. MIT Press, Cambridge.
[**5**] Quiroga RQ (2016) Magic and cognitive neuroscience. Current Biology 26(10):R387-407.
[**6**] Mathis A, Mamidanna P, Cury KM, Abe T, Murthy VN, Mathis MW, & Bethge M (2018) DeepLabCut: markerless pose estimation of user-defined body parts with deep learning. Nature Neuroscience 21: 1281–1289.
[**7**] Zhou Z & Firestone C (2019) Humans can decipher adversarial images. Nature Communications 10:1334.
[**8**] Pearl J & Mackenzie D (2018) The Book of Why: The New Science of Cause and Effect. Basic Books, NY.
[**9**] Lake BM, Ullman TD, Tenenbaum JB, Gershman SJ (2017) Building machines that learn and think like people. Behavioral and Brain Sciences 40:e253.
[**10**] Rahwan I et al. (2019) Machine behaviour. Nature 568, 477–486.

∗

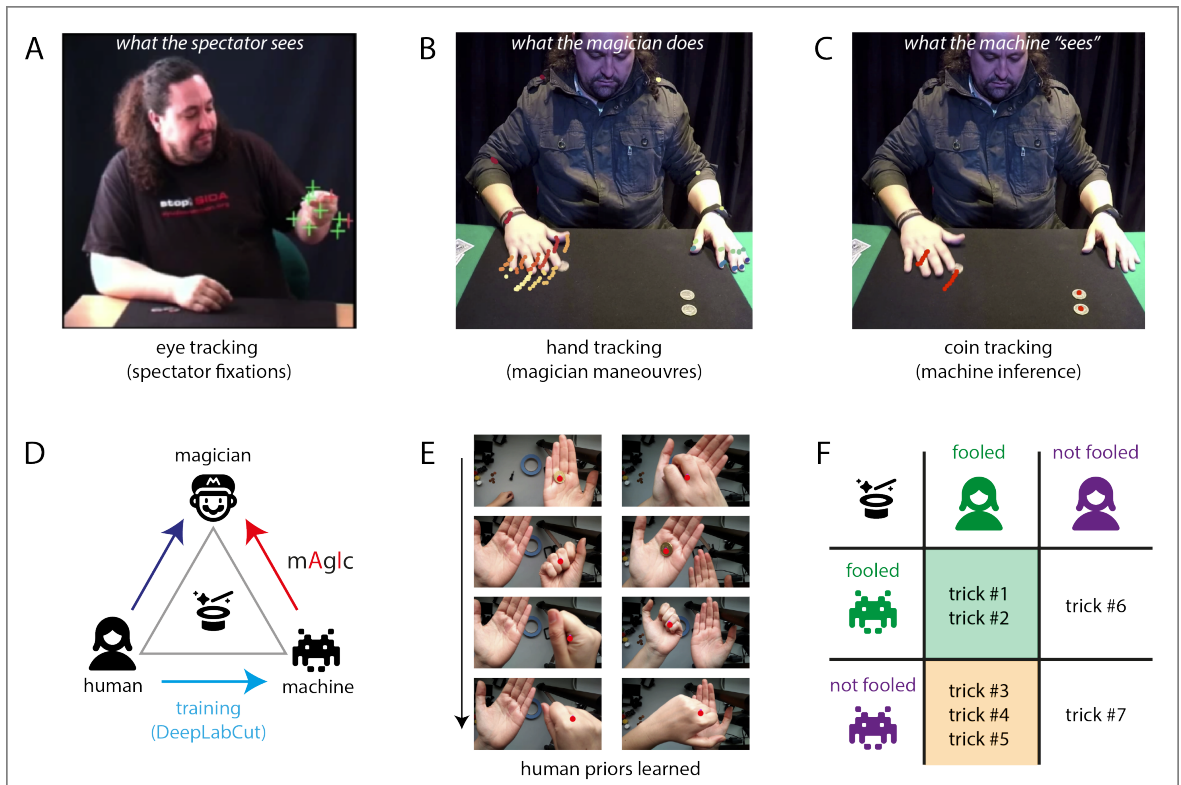

**Figure 1. Using AI in the context of magic to study human cognitive illusions.** (**A**) Eye tracking reveals where participants look at during a magic trick (adapted from [**5**]; donation from the author). (**B**) DeepLabCut [**6**] allows studying the magician's motor skills by measuring the precise location of all his fingers (nails and knuckles; and also wrists, elbows and shoulders) during a prestidigitation maneuver. (**C**) Tracking the position of four coins during a magic trick, and inferring them when not visible. (**D**) Rather than conceiving AI as a tracking tool to put "dots on spots", we imagine it is an "artificial spectator" watching the trick that the magician performs (red arrow). (**E**) Training the algorithm to follow visible coins and infer no visible ones (example frames, from top to bottom). (**F**) By deconstructing a magic show into simple maneuvers stripped from any verbal cues, we create brief purely motor tricks (see **Supplementary Video**; experience the magic yourself) and interpret whether they would count as magic for both a human and a machine (green quadrant) or only fool a human (orange quadrant). The lack of certain human biases in the algorithm can make us humans more aware of our own.